\documentclass{article}
\usepackage{spconf, amsmath, graphicx, multirow, bm, subfigure, enumitem, array}
\DeclareGraphicsExtensions{.pdf}
\graphicspath{{./Fig/}}

\title{Noise-Robust ASR for the third `CHiME' Challenge Exploiting \\ Time-Frequency Masking based Multi-Channel Speech Enhancement and \\ Recurrent Neural Network}
\name{
    Zaihu Pang, Fengyun Zhu
}
\address{
    Lingban Technology Co., Ltd. \\
    Beijing, China \\
    \{zhpang, fyzhu\}@ling-ban.com
}

\begin{document}
\maketitle
\begin{abstract}
%This paper proposes a noise-robust speech recognition system exploiting time-frequency masking based multi-channel speech enhancement and recurrent neural networks for the 3rd CHiME speech separation and recognition challenge.
In this paper, the Lingban entry to the third `CHiME' speech separation and recognition challenge is presented.
A time-frequency masking based speech enhancement front-end is proposed to suppress the environmental noise utilizing multi-channel coherence and spatial cues.
The state-of-the-art speech recognition techniques,
namely recurrent neural network based acoustic and language modeling,
state space minimum Bayes risk based discriminative acoustic modeling,
and i-vector based acoustic condition modeling,
are carefully integrated into the speech recognition back-end.
To further improve the system performance by fully exploiting the advantages of different technologies,
the final recognition results are obtained by lattice combination and rescoring.
Evaluations carried out on the official dataset prove the effectiveness of the proposed systems.
Comparing with the best baseline result, the proposed system obtains consistent improvements with over 57\% relative word error rate reduction on the real-data test set.
\end{abstract}
\section{Introduction}
\label{sec:intro}
Far-field noise-robust automatic speech recognition (ASR) in real-world environments is still a challenging problem.
The series of `CHiME' speech separation and recognition challenges offered a great opportunity
for the researchers from the signal processing community and the ASR community to work collaboratively toward this goal.
The past `CHiME' challenges have contributed to the development of several speech enhancement techniques,
and novel framework that integrate speech enhancement and ASR \cite{Barker2013, Vincent2013}.

%(
The main difference between the current `CHiME' challenge and the past ones is instead of working on simulated data only,
the current challenge focuses on real-world data: multi-channel recording using mobile tablet device in a variety of noisy public environments \cite{Barker2015}.
How to make use of both the real and simulated data to improve the system performance remains an open question for the current challenge.
%Since the speech enhancement methods often work better on simulated data, for systems with speech enhancement front-end, the influence of the mismatch between the two kinds of enhanced data should be examined.

In this paper, we presented the Lingban entry to the 3rd `CHiME' speech separation and recognition challenge.
A time-frequency masking based speech enhancement front-end is proposed to suppress the environmental noise utilizing multi-channel coherence and spatial cues.
An adaptive microphone array self-calibration method is adopted to overcome the problem of microphone mismatch.
%The state-of-the-art speech recognition techniques are adopted.
For acoustic modeling, neural network acoustic models including maxout neural network \cite{Swietojanski2014}
and long short term memory with project layers (LSTMP) \cite{Sak2014b, Li2015} are adopted.
Mel-frequency cepstral coefficient (MFCC) based feature-space maximum likelihood linear regression (fMLLR) features are utilized.
Moreover, online extracted i-vector is also used as the network input for encoding these effects: speaker, channel and background noise \cite{Dehak2011, Glembek2011, Saon2013, Senior2014}.
State-space Minimum Bayes Risk (sMBR) discriminative training is conducted on the neural networks based acoustic models \cite{Vesely2013, Sak2014a}.
For language and lexicon modeling, to model the inter-word silence more precisely, pronunciation lexicon with silence probability is adopted \cite{Chen2015}.
4-gram language model with Kneser-Ney smoothing is adopted in the first-pass decoding.
A language model based on jointly trained recurrent neural network and maximum entropy models (RNNME) is adopted for the second-pass rescoring \cite{Mikolov2012, Mikolov2011}.
Finally, to further improve the system performance by fully exploiting the advantages of different technologies,
the recognition results are obtained by lattice combination and rescoring.

Evaluations are carried out on the official dataset.
Comparing with the best baseline result, the proposed system obtains consistent improvements
with over 57\% and 42\% relative word error rate (WER) reduction on real and simulated test set respectively.

The rest of this paper is organized as follows.
%A brief overview of the system is described in Section 2.
The time-frequency masking based speech enhancement front-end is presented in Section 2.
The speech recognition back-end is presented in Section 3.
Experiments are presented in Section 4, followed by the conclusions in Section 5.

%\section{System overview}

\section{Speech enhancement}
\begin{figure*}[!htbp]
    \centering
    \includegraphics[width=6.8in]{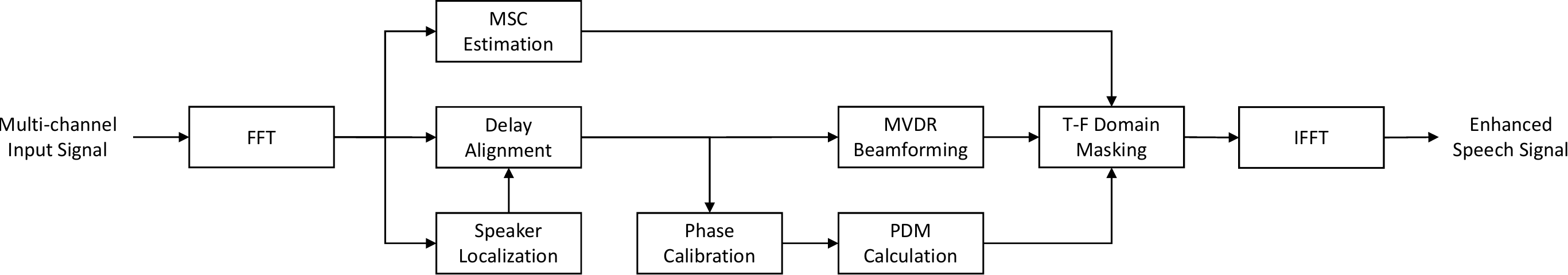}
    \caption{Schematic diagram of the proposed speech enhancement method.}
    \label{fig:diagram_enhancement}
\end{figure*}

Fig.\ref{fig:diagram_enhancement} shows a schematic diagram of the proposed speech enhancement method.
The proposed method is developed on the basis of the speech enhancement baseline of the `CHiME' challenge.
Keeping the analysis-synthesis scheme, the microphone failure detector, the SPR-PHAT based speaker localizer and the MVDR beamformer identical to the baseline,
in this study, time-frequency masking based on multi-channel magnitude-squared coherence (MSC) and phase difference measurement (PDM) is introduced.
The MSC-based masking is used to suppress the diffused noise, while the PDM-based masking is used to suppress the directional interference not coming from the target direction.
The time-frequency masking is performed by filtering the output subband signal of the MVDR beamformer with the said maskers.
Since the phase difference based methods are sensitive to the mismatch of microphone phase response, an adaptive microphone array self-calibration method is adopted.

\subsection{MSC based time-frequency masking}
The MSC between two signals $x(t)$ and $y(t)$ is defined as:
\begin{equation}
\label{eq:MSC}
C_{xy}(f) = \frac{\left|S_{xy}(f)\right|^2}{S_{xx}(f)S_{yy}(f)},
\end{equation}
where $f$ is the frequency, $S_{xy}(f)$ is the cross-spectral density between the two signals,
$S_{xx}(f)$ and $S_{yy}(f)$ are the auto-spectral density of $x$ and $y$ respectively.
The values of the coherence function, which always satisfy $0 \le C_{xy}(f) \le 1$, indicates the extent to which the power of $y$ could be predicted from $x$ by a linear system.
In multi-channel signal processing, the MSC is an efficient mean of noise reduction \cite{LeBouquin1992}.
In the case of microphone array with sufficiently large microphone spacing,
incoherent (diffused) noise would be indicated by small coherence values,
while the directional signals would have large coherence values \cite{Brandstein2001, Trawicki2012}.

In this study, MSC between all the microphone pairs (except the channels detected to be failed) are estimated by Welch's method of periodogram.
The MSC-based time-frequency masker is derived by averaging the MSC over all the microphone pairs.

\subsection{PDM based time-frequency masking}
In the past `CHiME' challenges, phase difference based time-frequency masking noise suppression front-ends was proved to be effective in the case of dual-channel simulated data \cite{Tachioka2013}.
After steering the multi-channel signals towards the target direction by delay alignment, inter-channel phase differences could be obtained.
Time-frequency bins which has a phase difference not close to zero are less likely to be the ones from the target direction.

In the current `CHiME' challenge, the experimental condition is extended from dual-channel simulated data to 6-channel real-world recordings.
For the multi-channel subband input signals, delay alignment is performed by steering the signals towards the target direction given by the speaker localizer.
Phase differences between all the microphone pairs (except the 2nd channel and the failed ones) could be calculated for all the frequency bins.
The PDM could be obtained by averaging the absolute value of the phase differences over all the microphone pairs.
Assuming that the estimated speaker location is correct, the bins dominated by the signal coming from the target direction would have small PDM values.

Since the PDM values don't lies within the range of $[0, 1]$,
the PDM based time-frequency masker $W'_{P}(f, t)$ is obtained by performing a non-linear transformation on the PDM value $W_{P}(f, t)$,
and hard clipped to have a maximum value of $1$:
\begin{equation}
\label{eq:MSC}
W'_{P}(f, t) = clip(1 - tanh(W_{P}(f, t) - \alpha(f))),
\end{equation}
where $\alpha(f)$ is a bias function, which is introduced to fit the frequency-dependent distribution of the PDM values.
$\alpha(f)$ is empirically determined by dividing the PDM values in the training set into signal and noise, prior to the testing stage.
In this study, $\alpha(f) = 0.4 + 0.3f / f_s$, where $f_s$ is the sampling rate, is used.

\subsection{Microphone array self-calibration}
It's well known that microphone mismatch could cause speech signal cancellation in adaptive beamforming \cite{Haykin2010}.
Similarly, in the context of phase difference based time-frequency masking, the accuracy of the masking function would be degraded by the interference of
the unknown phase response of the microphones, and the environment-dependent phase response of the transfer function between the speaker and the microphones.

In this study, adaptive microphone array self-calibration with recursive configuration is adopted \cite{Buck2006}.
For each channel, a transfer function is estimated by minimizing the error between the output signal of a delay-and-sum beamformer and the delay aligned microphone signal.
To evaluate the effectiveness of the self-calibration in the context of phase difference base time-frequency masking, only the phase response of the transfer function is used, leaving the MVDR beamformer identical to the baseline.

In this study, a 2-stage self-calibration scheme is proposed.
In the first stage, an off-line calibration is done on the training set,
which provides a robust estimation of the microphone phase response,
due to the relatively large amount of training data.
The parameters of the calibration filters are updated in a batched fashion, using the frames with SNR greater than the utterance-level median SNR.
The SNR is estimated using the method provided by the acoustic simulation baseline.
In the second stage, after the phase response of the first calibration filter is compensated,
an on-line calibration is done for each utterance in the test set,
which provides an estimation of the environment-dependent transfer function.
The parameters of the calibration filters are updated in a batched fashion, using all the frames in the evaluation utterance.
The phase responses of the two calibration filters are compensated before the PDM calculation.

\section{Speech recognition}

%\subsection{Acoustic modeling}
%First, neural network acoustic models including maxout neural network \cite{} and long short term memory with project layers (LSTMP) \cite{Sak2014b, Li2015} are adopted.
%Second, Mel-frequency cepstral coefficient (MFCC) based feature-space maximum likelihood linear regression (fMLLR) features are utilized.
%Moreover, since the noise environment condition is unknown, utterance-level online i-vector feature is adopted for encoding the acoustic conditions \cite{Dehak2011, Glembek2011, Saon2013, Senior2014}.
%Third, state-space minimum Bayes risk (sMBR) discriminative training is conducted on the neural networks based acoustic models \cite{Vesely2013, Sak2014a}.
%Finally, system combination is introduced to further improve the system performance by integrating systems with different properties using lattice combination.
%
%\subsection{Language modeling}
%First, to model the inter-word silence more precisely, pronunciation lexicon with silence probability is adopted \cite{Chen2015}.
%Second, the baseline 3-gram language model is replaced by a 4-gram language model with Kneser-Ney smoothing trained on the official training set using the SRILM toolkit \cite{Stolcke2002}.
%Third, an RNNME based language model is adopted for the second-pass rescoring \cite{Mikolov2012, Mikolov2011}.

In recent years, the recurrent neural networks (RNNs) based speech recognition systems have brought about noteworthy performance improvements.
Specifically for acoustic modeling, the long short-term memory (LSTM) based deep networks have been shown to give the state-of-the-art performance on some of the speech recognition tasks.
In the seminal work, Graves \textit{et al.} \cite{Graves2013} proposed to use stacked bidirectional LSTM trained with connectionist temporal classification \cite{Graves2006} for phoneme recognition.
Subsequently, LSTM RNNs have been successfully applied and shown to give state-of-the-art performance on robust speech recognition task \cite{Geiger2014},
and many large vocabulary speech recognition tasks \cite{Sak2014b, Li2015}.
On the other hand, RNN based language models have also shown advantages over the conventional N-gram language models in both perplexity and speech recognition error rate.

The speaker variability is also an important issue.
Efforts have been made to train acoustic models using speaker-adapted features,
which can be obtained by speaker normalization techniques such as the vocal tract length normalization (VTLN) and fMLLR \cite{Seide2011}.
Another group of approaches are using speaker-aware training \cite{Yu2015},
where the speaker information, such as speaker code \cite{Abdel2013} and speaker- or utterance-level i-vector \cite{Saon2013, Senior2014}, is provided to the neural networks directly.
It should be noticed that the information about channel and background noise is also encoded by the i-vector.

In this study, Mel-scale log-filterbank coefficients (FBANK) features, MFCC-based fMLLR features are used.
The i-vector was extracted online and concatenated to the acoustic feature.
The hybrid approach for acoustic modeling is adopted,
in which the neural networks’ outputs are converted to pseudo likelihood,
and used as the state output probability of the HMMs.
The networks are trained based on alignments from GMM-HMM systems.
Neural network acoustic models including maxout neural network \cite{Swietojanski2014} and LSTMP \cite{Sak2014b, Li2015} are adopted.
sMBR discriminative training is conducted on the neural networks based acoustic models \cite{Vesely2013, Sak2014a}.
To model the inter-word silence more precisely, pronunciation lexicon with silence probability is adopted \cite{Chen2015}.
The baseline 3-gram language model is replaced by a 4-gram language model with Kneser-Ney smoothing trained on the official training set.
An RNNME based language model \cite{Mikolov2012, Mikolov2011} is adopted for the second-pass rescoring.
Finally, system combination is introduced to further improve the system performance by integrating systems with different properties using lattice combination and rescoring.

\section{Experiments and discussions}
\subsection{Experimental setups}
All the experiments were conducted on the official dataset.
The GMM-HMM system was trained using Kaldi \cite{Povey2011}.
All the networks were trained based on the alignment result of the GMM-HMM system with 2824 tied context dependent HMM states.
In the training procedure of LSTM networks, the strategy introduced by \cite{Pascanu2012} was applied to scale down the gradients.
Besides, since the information from the future frames helps the LSTM networks making better decisions, we also delayed the output HMM state labels by 4 frames.
The feed-forward DNNs used the concatenated features, which were produced by concatenating the current frame with 7 frames in its left and right context.
The inputs to LSTMP networks were only the current frames.

\subsection{Speech enhancement}
%The results show that the MSC and PDM based speech enhancement could improve the system performance significantly.
%客观描述开发集和测试集上提出方法在real和simulated两个测试集上的性能提升情况（绝对和相对）。
%详细讨论在real和simulated提升差异的原因（即讨论和分析在simulated上提升有限的原因）。
%分析系统提出方法在实际应用中的意义。
%其它结果做同样的描述、讨论、分析。
\begin{table}[!t]
\small
\renewcommand{\arraystretch}{1.1}
\caption{Evaluation results (WER[\%]) of the proposed speech enhancement front-ends with the baseline acoustic model (GMM-HMM) and language model (3-gram).
``CAL'' means phase calibration.}
\label{table_exp_enh}
\centering
\begin{tabular}{|c|p{25pt}<{\centering}|p{25pt}<{\centering}|p{25pt}<{\centering}|p{25pt}<{\centering}|}
    \hline
    \multirow{2}{*}{Speech Enh.} &
    \multicolumn{2}{c|}{Dev. set} &
    \multicolumn{2}{c|}{Test set} \\
    \cline{2-5}
    {}                      & {Real}  & {Simu.} & {Real}            & {Simu.} \\ \hline
    {Baseline (noisy)}      & {18.70} & {18.71} & {33.23}           & {21.59} \\ \hline
    {Baseline (enhanced)}   & {20.55} & { 9.79} & {37.36}           & {10.59} \\ \hline \hline
    {MSC}                   & {15.12} & { 9.57} & {26.42}           & {12.01} \\ \hline
    {PDM}                   & {13.26} & { 8.23} & {26.23}           & {10.20} \\ \hline
    {MSC+PDM}               & {12.45} & { 8.45} & \bf{24.72}        & {11.27} \\ \hline \hline
    {PDM+CAL}               & {13.07} & { 7.96} & {25.11}           & {10.33} \\ \hline
    {MSC+PDM+CAL}           & {12.14} & { 8.71} & \bf{24.55}        & {11.51} \\ \hline
    \end{tabular}
\end{table}

Table \ref{table_exp_enh} shows the evaluation results of the proposed speech enhancement methods with the baseline acoustic model (GMM-HMM) and language model (3-gram).
Comparing with the baseline using noisy data,
the performance of the baseline with speech enhancement front-end is greatly improved on simulated test set,
but degraded on real test set.
This phenomenon shows that mismatch between enhanced real and simulated data is introduced by the baseline beamforming based speech enhancement process.
The system performance on the real test set would be greatly influenced by this mismatch, since relatively large amount of simulated data is used in the training phase.

By introducing the MSC based time-frequency masking into the baseline speech enhancement front-end,
the WER on real test set is reduced by 10.94\%.
It proves the effectiveness of the MSC based time-frequency masking.
By contrast, the WER on simulated test set is increased by 1.42\%.
It shows that this method performs more equally on real and simulated data.
Hence the said mismatch is weakened.
The effectiveness of the PDM based time-frequency masking is also proved by the WER reduction on real test set of 11.13\%.
While the WER on simulated test set is lower than the MSC based system.
Comparing to the MSC based method,
although better performance on real test set is achieved,
greater mismatch between real and simulated data is introduced by the PDM based method.
By using the two kinds of time-frequency masking together, an absolute WER reduction of 12.64\% is achieved.

It's also shown that the by introducing phase calibration, further absolute WER reduction of 1.12\% and 0.17\% is made on the basis of the ``PDM'' and ``MSC+PDM'' systems respectively.
The mismatch between real and simulated data is further weakened.
It should be noticed that, since the time for the challenge is limited, further experiments for the phase calibration is \textit{not} conducted.
In the following experiments, the ``MSC+PDM'' front-end is used.

\subsection{Lexicon and language modeling}
Table \ref{table_exp_lm} shows the evaluation results of different language models with the proposed front-end (MSC+PDM) and the baseline acoustic model (GMM-HMM).
The N-gram language models are used in the first-pass decoding, while the RNNME-based language model is used for the second-pass rescoring.

\begin{table}[!t]
\small
\renewcommand{\arraystretch}{1.1}
\caption{Evaluation results (WER[\%]) of different language models with the proposed front-end (MSC+PDM) and the baseline acoustic model (GMM-HMM).
    ``SP'' means pronunciation lexicon with silence probability. ``RNNME'' means RNNME-based language model.}
\label{table_exp_lm}
\centering
\begin{tabular}{|c|p{25pt}<{\centering}|p{25pt}<{\centering}|p{25pt}<{\centering}|p{25pt}<{\centering}|}
    \hline
    \multirow{2}{*}{Language model} &
    \multicolumn{2}{c|}{Dev. set} &
    \multicolumn{2}{c|}{Test set} \\
    \cline{2-5}
    {}    & {Real} & {Simu.} & {Real} & {Simu.} \\ \hline
    {Baseline (3-gram)}     & {12.45} & {8.45} & {24.72}       & {11.27} \\ \hline
    {3-gram+SP}             & {11.70} & {8.23} & {23.79}       & {11.11} \\ \hline
    {4-gram+SP}             & { 9.93} & {7.59} & \bf{20.42}    & {9.96} \\ \hline
    {4-gram+SP+RNNME}       & { 9.38} & {7.05} & \bf{19.68}    & {9.25} \\ \hline
    \end{tabular}
\end{table}

Experiment results show that the system performance is improved consistently by introducing pronunciation lexicon with silence probability.
The system performance is further improved by introducing the 4-gram language model and RNNME-based language model.

\subsection{Acoustic modeling}
Table \ref{table_exp_am} shows the evaluation results of different features and different acoustic models with the proposed front-end (MSC+PDM) and language model (4-gram+SP).
It should be noted that the RNNME based second-pass rescoring is not used in this evaluation.

The maxout deep neural network has 4 hidden layers and each layer has 1000 neurons with a group size of 4.
The LSTMP recurrent neural network has a single hidden layer with 1000 neurons and 700 projection units.
The 40-dimensional fMLLR features are attained based on the 13-dimensional MFCC features.
The fMLLR transformation matrix is estimated using the baseline GMM-HMM.
The 50-dimensional i-vector features are extracted on utterance level.
The language model used in the sMBR discriminative training is the baseline 3-gram language model.

\begin{table}[!t]
\small
\renewcommand{\arraystretch}{1.1}
\caption{Evaluation results (WER[\%]) of different features and acoustic models with the proposed front-end (MSC+PDM) and language model (4-gram+SP).
    ``dp'' means dropout.
    ``fM'' and ``iV'' in the lower part of the table means fMLLR and i-vector for short.
}
\label{table_exp_am}
\centering
\begin{tabular}{|c|c|p{17pt}<{\centering}|p{17pt}<{\centering}|p{17pt}<{\centering}|p{17pt}<{\centering}|}
    \hline
    \multirow{2}{*}{Feature} &
    \multirow{2}{*}{Model} &
    \multicolumn{2}{c|}{Dev. set} &
    \multicolumn{2}{c|}{Test set} \\
    \cline{3-6}
    &                               & {Real}        & {Simu.}      & {Real}        & {Simu.}      \\ \hline
    {FBank}     & Maxout            & {10.96}       & {7.75}       & {21.33}       & {9.68}       \\ \hline
    {FBank}     & LSTMP             & {10.96}       & {7.02}       & {20.82}       & {9.21}       \\ \hline \hline
    {fMLLR}     & Maxout            & { 9.44}       & {7.08}       & {18.41}       & {8.57}       \\ \hline
    {fMLLR}     & Maxout+dp         & { 9.37}       & {6.86}       & {18.55}       & {8.59}       \\ \hline
    {fMLLR}     & LSTMP             & { 8.88}       & {6.07}       & \bf{16.86}    & {7.56}       \\ \hline \hline
    {fM+iV}     & Maxout+dp         & { 9.37}       & {6.89}       & {18.42}       & {8.61}       \\ \hline
    {fM+iV}     & LSTMP             & { 8.32}       & {5.86}       & \bf{16.45}    & {7.20}       \\ \hline \hline
    {fM+iV}     & Maxout+dp+sMBR    & { 8.71}       & {6.42}       & {17.60}       & {8.09}       \\ \hline
    {fM+iV}     & LSTMP+sMBR        & { 7.91}       & {5.43}       & \bf{15.55}    & {6.86}       \\ \hline
    \end{tabular}						
\end{table}												
%fM 建议也用fMLLR,把前两列合并（用 / 隔开即可），表格能展示

%1、特征相同时（FBank、fMLLR），声学建模方法LSTMP性能比Maxout好。
%2、声学建模方法相同时，fMLLR比FBank性能好。
%3、在fMLLR特征基础上，增加ivector特征，性能得到进一步提升。
%4、引入区分性训练方法sMBR,性能进一步提升，LSTM+sMBR性能优于Maxout+sMBR.
Experimental result shows that the LSTMP based method performs better than the maxout based method while using the same features.
While using same acoustic modeling methods, fMLLR based feature performs better than Fbank based feature.
By combining the fMLLR based feature with the i-vector based feature, the system performance improved consistently.
By using sMBR based discriminative training, the system performance improved consistently.
The best WER of 15.55\% is achieved by the ``LSTMP+sMBR'' on real data, and the second best acoustic modeling method is ``Maxout+dp+sMBR''.

%\subsection{System combination}
Table \ref{table_exp_cmb} shows the evaluation results of the combined systems.
Since the RNNME based second-pass rescoring is not included in the evaluation of acoustic modeling,
the RNNME based second-pass rescoring is introduced into the two best-performed acoustic models (Maxout+dp+sMBR, LSTMP+sMBR) with the 4-gram+SP language model in one-pass decoding.
Moreover, the lattice combination is performed on the two intergraded systems to achieve a better performance.
The front-end used in this evaluation is MSC+PDM.
The feature used in this evaluation is fMLLR+i-vector.

\begin{table}[!t]
\small
\renewcommand{\arraystretch}{1.1}
\caption{
    Evaluation results (WER[\%]) of the combined systems.
}
\label{table_exp_cmb}
\centering
\begin{tabular}{|c|p{17pt}<{\centering}|p{17pt}<{\centering}|p{17pt}<{\centering}|p{17pt}<{\centering}|}
    \hline
    \multirow{2}{*}{System} &
    \multicolumn{2}{c|}{Dev. set} &
    \multicolumn{2}{c|}{Test set} \\
    \cline{2-5}
    & {Real} & {Simu.} & {Real} & {Simu.} \\ \hline
    {Maxout+dp+sMBR,} & \multirow{2}{*}{8.06} & \multirow{2}{*}{5.90} & \multirow{2}{*}{16.93} & \multirow{2}{*}{7.59} \\
    {4-gram+SP, RNNME}   & & & & \\ \hline
    {LSTMP+sMBR,}     & \multirow{2}{*}{7.20} & \multirow{2}{*}{4.85} & \multirow{2}{*}{14.69} & \multirow{2}{*}{6.27} \\
    {4-gram+SP, RNNME}   & & & & \\ \hline
    {Maxout+dp+sMBR,} & \multirow{3}{*}{7.06} & \multirow{3}{*}{4.86} & \multirow{3}{*}{\bf{14.28}} & \multirow{3}{*}{6.14} \\
    {LSTMP+sMBR,}     & & & & \\
    {4-gram+SP, RNNME}   & & & & \\ \hline
    \end{tabular}
\end{table}

%1、实验结果显示使用最好的语言模型4-gram,且进行RNNME rescoring，第一第二最优的声学模型的性能都得到了进一步的提升。
%2、同样，还是LSTMP+sMBR性能更佳。
%3、两个系统词网结果合并后，识别性能得到进一步提升。
Experimental result shows that the ``LSTMP+sMBR'' and ``Maxout+dp+sMBR'' systems obtain further improvement by introducing RNNME-based language model second-pass rescoring.
Through lattice combination, a final WER of 14.28\% is achieved on the real test set.
%4、讨论两个声学模型性能贡献。参考文献！！！

\subsection{Overall comparison}
Table \ref{table_exp_ove} shows the results of the overall comparison of the proposed systems and the baseline systems:
\begin{itemize}
\setlength{\itemsep}{0pt}
\setlength{\parsep}{0pt}
\setlength{\parskip}{0pt}
\item Baseline I: the GMM-HMM baseline on noisy data;
\item Baseline II: the GMM-HMM baseline on enhanced data;
\item System I: proposed speech enhancement front-end with baseline acoustic model (GMM-HMM) and language model (3-gram);
\item System II: proposed language model with baseline acoustic model (GMM-HMM) on noisy data;
\item System III: proposed acoustic model with baseline language model (3-gram) on noisy data;
\item System IV: the best-performed proposed system.
\end{itemize}
Table \ref{table_exp_env} shows the WERs for the 4 acoustic environments achieved by the best system.

\begin{table}[!t]
\small
\renewcommand{\arraystretch}{1.1}
\caption{Overall comparison of the evaluation results (WER[\%]) of the proposed systems and the baseline systems.}
\label{table_exp_ove}
\centering
\begin{tabular}{|c|p{25pt}<{\centering}|p{25pt}<{\centering}|p{25pt}<{\centering}|p{25pt}<{\centering}|}
    \hline
    \multirow{2}{*}{System} &
    \multicolumn{2}{c|}{Development set} &
    \multicolumn{2}{c|}{Test set} \\
    \cline{2-5}
    {}            & {Real}  & {Simu.} & {Real}     & {Simu.} \\ \hline
    {Baseline I}  & {18.70} & {18.71} & {33.23}    & {21.59} \\ \hline
    {Baseline II} & {20.55} & { 9.79} & {37.36}    & {10.59} \\ \hline \hline
    {System I}    & {12.14} & { 8.71} & {24.55}    & {11.51} \\ \hline
    {System II}   & {15.55} & {15.49} & {28.02}    & {18.26} \\ \hline
    {System III}  & {12.38} & {10.91} & {21.47}    & {14.09} \\ \hline \hline
    {System IV}   & { 7.06} & { 4.86} & \bf{14.28} & { 6.14} \\ \hline
    \end{tabular}			
\end{table}						

%单独使用任何一项技术（信号处理、语言模型、声学模型），性能都得到大幅度提升。
%综合利用三项技术，性能提升更大。
Experimental results from the upper part of table \ref{table_exp_ove} show the contribution of the proposed front-end, acoustic model and language model separately.
Comparing to the baseline system on enhanced data,
by introducing the proposed speech enhancement method, an absolute WER reduction of 12.81\% is achieved.
Comparing to the baseline systems using noisy data,
absolute WER reduction of 11.76\% and 5.21\% is achieved by the proposed acoustic model and language model respectively.
The system IV, which make use of the proposed front-end and back-end, performs the best result.
A final WER of 14.28\% is achieved on the real test set.

\begin{table}[!t]
\small
\renewcommand{\arraystretch}{1.1}
\caption{WERs[\%] for the 4 acoustic environments achieved by the best system.}
\label{table_exp_env}
\centering
\begin{tabular}{|c|p{30pt}<{\centering}|p{30pt}<{\centering}|p{30pt}<{\centering}|p{30pt}<{\centering}|}
    \hline
    \multirow{2}{*}{Environment} &
    \multicolumn{2}{c|}{Development set} &
    \multicolumn{2}{c|}{Test set} \\
    \cline{2-5}
    {}    & {Real} & {Simu.} & {Real} & {Simu.} \\ \hline
    {BUS} & {8.87} & {4.01} & {16.19} & {4.45} \\ \hline
    {CAF} & {5.50} & {6.05} & {13.37} & {6.48} \\ \hline
    {PED} & {5.69} & {4.25} & {17.02} & {6.05} \\ \hline
    {STR} & {8.17} & {5.12} & {10.53} & {7.58} \\ \hline
    \end{tabular}
\end{table}

%四种场景下的性能比较，测试集上(test real)来看 STR性能最好，而PED最差。
%与devlopment集，不一致，分析原因，不同场景下利用的信息不一致，或者噪音的平稳性不一致？
%real 与 simu 不一致，分析原因，应该是speech enhancement 的困难所在。
%分析该任务在日用场景下的难点与重点，下一步重点需要攻克的问题。

\section{Conclusions}
\label{sec:ref}
In this paper, we presented the Lingban entry to the 3rd `CHiME' speech separation and recognition challenge.
A time-frequency masking based speech enhancement front-end is proposed.
The state-of-the-art recurrent neural networks based acoustic and language modeling methods are adopted.
Evaluations are carried out on the official dataset.
Comparing with the best baseline result, the proposed system obtains consistent improvements with over 57\% relative WER reduction.

Since the front-end and back-end are working separately in the current implementation,
a unified end-to-end learning framework for multi-channel noise-robust ASR would be proposed in the future works.
Furthermore, the proposed methods should be evaluated on real-application tasks, such as spontaneous speech recognition using general purpose commercial mobile devices with less microphones.

\section{ACKNOWLEDGMENT}
The authors would like to thank Zhiping Zhang, Xiangang Li, Yi Liu and Tong Fu for their kindly helps.

\bibliographystyle{IEEEbib}
\bibliography{strings,refs}

\begin{thebibliography}{10}

\bibitem{Barker2013}
J.~Barker, E.~Vincent, N.~Ma, H.~Christensen, and P.~Green,
\newblock ``The {PASCAL} {CHiME} speech separation and recognition challenge,''
\newblock {\em Computer Speech \& Language}, vol. 27, no. 3, pp. 621--633, May
  2013.

\bibitem{Vincent2013}
E.~Vincent, J.~Barker, S.~Watanabe, J.~Le~Roux, F.~Nesta, and M.~Matassoni,
\newblock ``The second {CHiME} speech separation and recognition challenge an
  overview of challenge systems and outcomes,''
\newblock in {\em Proceedings of the 2013 {IEEE} {Workshop} on {Automatic}
  {Speech} {Recognition} and {Understanding} ({ASRU})}, 2013.

\bibitem{Barker2015}
J.~Barker, R.~Marxer, E.~Vincent, and S.~Watanabe,
\newblock ``The third `{CHiME}' speech separation and recognition challenge:
  Dataset, task and baselines,''
\newblock in {\em Proceedings of the 2015 IEEE Workshop on Automatic Speech
  Recognition and Understanding (ASRU)}, 2015.

\bibitem{Swietojanski2014}
P.~Swietojanski, J.~Li, and J.~Huang,
\newblock ``Investigation of maxout networks for speech recognition,''
\newblock in {\em Proceedings of the 2014 {ICASSP}}, 2014, pp. 7649--7653.

\bibitem{Sak2014b}
H.~Sak, A.~Senior, and F.~Beaufays,
\newblock ``Long short-term memory recurrent neural network architectures for
  large scale acoustic modeling,''
\newblock in {\em Proceedings of the 2014 {INTERSPEECH}}, 2014, pp. 338--342.

\bibitem{Li2015}
X.~Li and X.~Wu,
\newblock ``Constructing long short-term memory based deep recurrent neural
  network for large vocabulary speech recognition,''
\newblock in {\em Proceedings of the 2015 {ICASSP}}, 2015.

\bibitem{Dehak2011}
N.~Dehak, P.~Kenny, R.~Dehak, P.~Dumouchel, and P.~Ouellet,
\newblock ``Front-end factor analysis for speaker verification,''
\newblock {\em IEEE Transactions on Audio, Speech, and Language Processing},
  vol. 19, no. 4, pp. 788--798, 2011.

\bibitem{Glembek2011}
O.~Glembek, L.~Burget, P.~Matejka, M.~Karafi{\'a}t, and P.~Kenny,
\newblock ``Simplification and optimization of i-vector extraction,''
\newblock in {\em Proceedings of the 2011 {ICASSP}}, 2011, pp. 4516--4519.

\bibitem{Saon2013}
G.~Saon, H.~Soltau, D.~Nahamoo, and M.~Picheny,
\newblock ``Speaker adaptation of neural network acoustic models using
  i-vectors,''
\newblock in {\em Proceedings of the 2013 IEEE Workshop on Automatic Speech
  Recognition and Understanding (ASRU)}, 2013, pp. 55--59.

\bibitem{Senior2014}
A.~Senior and I.~Lopez-Moreno,
\newblock ``Improving dnn speaker independence with i-vector inputs,''
\newblock in {\em Proceedings of the 2014 {ICASSP}}, 2014, pp. 225--229.

\bibitem{Vesely2013}
K.~Vesel{\'y}, A.~Ghoshal, L.~Burget, and D.~Povey,
\newblock ``Sequence-discriminative training of deep neural networks,''
\newblock in {\em Proceedings of the 2013 {INTERSPEECH}}, 2013.

\bibitem{Sak2014a}
H.~Sak, O.~Vinyals, G.~Heigold, A.~Senior, E.~McDermott, R.~Monga, and M.~Mao,
\newblock ``Sequence discriminative distributed training of long short-term
  memory recurrent neural networks,''
\newblock in {\em Proceedings of the 2014 {INTERSPEECH}}, 2014, pp. 1209--1213.

\bibitem{Chen2015}
G.~Chen, H.~Xu, M.~Wu, D.~Povey, and S.~Khudanpur,
\newblock ``Pronunciation and silence probability modeling for {ASR},''
\newblock in {\em Proceedings of the 2015 {INTERSPEECH}}, 2015.

\bibitem{Mikolov2012}
T.~Mikolov,
\newblock {\em Statistical language models based on neural networks},
\newblock Ph.D. thesis, Brno University of Technology, 2012.

\bibitem{Mikolov2011}
T.~Mikolov, S.~Kombrink, A.~Deoras, L.~Burget, and J.~\v{C}ernock\'{y},
\newblock ``{RNNLM} - recurrent neural network language modeling toolkit,''
\newblock in {\em Proceedings of the 2011 IEEE Workshop on Automatic Speech
  Recognition and Understanding (ASRU)}, 2011, pp. 196--201.

\bibitem{LeBouquin1992}
R.~Le~Bouquin and G.~Faucon,
\newblock ``Using the coherence function for noise reduction,''
\newblock {\em Communications, Speech and Vision, IEE Proceedings I}, vol. 139,
  no. 3, pp. 276--280, June 1992.

\bibitem{Brandstein2001}
M.~Brandstein and D.~Ward, Eds.,
\newblock {\em Microphone {Arrays}},
\newblock Springer-Verlag, New York, NY, 2001.

\bibitem{Trawicki2012}
M.~B. Trawicki and M.~T. Johnson,
\newblock ``Distributed multichannel speech enhancement with minimum
  mean-square error short-time spectral amplitude, log-spectral amplitude, and
  spectral phase estimation,''
\newblock {\em Signal Processing}, vol. 92, no. 2, pp. 345--356, Feb. 2012.

\bibitem{Tachioka2013}
Y.~Tachioka, S.~Watanabe, J.~Le~Roux, and J.~R. Hershey,
\newblock ``Discriminative methods for noise robust speech recognition: {A}
  {CHiME} {Challenge} {Benchmark},''
\newblock in {\em The 2nd {CHiME} {Workshop} on {Machine} {Listening} in
  {Multisource} {Environments}}, Vancouver, Canada, June 2013, pp. 19--24.

\bibitem{Haykin2010}
S.~Haykin and K.~J.~R. Liu,
\newblock {\em Handbook on {Array} {Processing} and {Sensor} {Networks}},
\newblock Wiley-IEEE Press, Jan. 2010.

\bibitem{Buck2006}
M.~Buck, T.~Haulick, and H.~Pfleiderer,
\newblock ``Self-calibrating microphone arrays for speech signal acquisition:
  {A} systematic approach,''
\newblock {\em Signal Processing}, vol. 86, no. 6, pp. 1230--1238, June 2006.

\bibitem{Graves2013}
A.~Graves, A.~Mohamed, and G.~Hinton,
\newblock ``Speech recognition with deep recurrent neural networks,''
\newblock in {\em Proceedings of the 2013 {ICASSP}}, 2013, pp. 6645--6649.

\bibitem{Graves2006}
A.~Graves, S.~Fern{\'a}ndez, F.~Gomez, and J.~Schmidhuber,
\newblock ``Connectionist temporal classification: Labelling unsegmented
  sequence data with recurrent neural networks,''
\newblock in {\em Proceedings of the 23rd International Conference on Machine
  Learning (ICML)}, 2006, pp. 369--376.

\bibitem{Geiger2014}
J.~Geiger, Z.~Zhang, F.~Weninger, B.~Schuller, and G.~Rigoll,
\newblock ``Robust speech recognition using long short-term memory recurrent
  neural networks for hybrid acoustic modelling,''
\newblock in {\em Proceedings of the 2014 {INTERSPEECH}}, 2014, pp. 631--635.

\bibitem{Seide2011}
F.~Seide, G.~Li, X.~Chen, and D.~Yu,
\newblock ``Feature engineering in context-dependent deep neural networks for
  conversational speech transcription,''
\newblock in {\em Proceedings of the 2011 IEEE Workshop on Automatic Speech
  Recognition and Understanding (ASRU)}, 2011, pp. 24--29.

\bibitem{Yu2015}
D.~Yu and L.~Deng,
\newblock {\em Automatic speech recognition - A deep learning approach},
\newblock Springer-Verlag London, 2015.

\bibitem{Abdel2013}
O.~Abdel-Hamid and H.~Jiang,
\newblock ``Fast speaker adaptation of hybrid nn/hmm model for speech
  recognition based on discriminative learning of speaker code,''
\newblock in {\em Proceedings of the 2013 {ICASSP}}, 2013, pp. 7942--7946.

\bibitem{Povey2011}
D.~Povey, A.~Ghoshal, G.~Boulianne, L.~Burget, O.~Glembek, N.~Goel,
  M.~Hannemann, P.~Motl{\'\i}{\v{c}}ek, Y.~Qian, P.~Schwarz, J.~Silovsk{\'y},
  G.~Stemmer, and K.~Vesel{\'y},
\newblock ``The {Kaldi} speech recognition toolkit,''
\newblock in {\em Proceedings of the 2011 IEEE Workshop on Automatic Speech
  Recognition and Understanding (ASRU)}, 2011.

\bibitem{Pascanu2012}
R.~Pascanu, T.~Mikolov, and Y.~Bengio,
\newblock ``On the difficulty of training recurrent neural networks,''
\newblock {\em CoRR}, vol. 1211.5063, 2012.

\end{thebibliography}

\end{document}